\documentclass{phc}
\usepackage{graphicx,amssymb}

\begin{document}
\begin{frontmatter}

\title{ Mechanism of the reorientation of stripes
        in the cuprates }
	
\author{      Marcin Raczkowski$^{a,b}$ }
\author{      Raymond Fr\'esard$^{b}$ }
\author{      Andrzej M. Ole\'{s}$^{a,c}$ }

\corauth[cor]{Corresponding author; e-mail: a.m.oles@fkf.mpg.de. }

\address{$^a$Marian Smoluchowski Institute of Physics, Jagellonian
              University, Reymonta 4, PL-30059 Krak\'ow, Poland }
\address{$^b$Laboratoire CRISMAT, UMR CNRS--ENSICAEN(ISMRA) 6508, 
              F-14050 Caen, France }
\address{$^c$Max-Planck-Institut f\"ur Festk\"orperforschung,
              Heisenbergstrasse 1, D-70569 Stuttgart, Germany }

\begin{abstract}
Using the mean field theory in the slave-boson approach we analyzed 
the electron correlation effects in the stripe phases. One finds that 
a finite next-nearest neighbor hopping $t'$ plays an important role in 
the low doping regime, where it controls the crossover from the filled 
diagonal to half-filled vertical/horizontal stripes at doping $
x\simeq 1/16$. 
\end{abstract}
\end{frontmatter}

\newpage

The dependence of magnetic correlations on the doping in high 
temperature superconductors is the subject of intense recent 
experimental studies \cite{Kas98}. They confirmed the theoretical 
prediction of self-organized one-dimensional structures called stripe 
phases \cite{Zaa89}, which act as antiphase domain walls for the 
antiferromagnetic (AF) order in the doped cuprates. Understanding of 
the stability of the half-filled vertical stripe (HVS) metallic phase
versus filled diagonal stripe (FDS) insulating phase requires the
analysis of electron correlation effects \cite{Fle01,Sei04,Rac06}. 
Here we analyze the microscopic origin of the observed reorientation
from the FDS phase to the HVS one, observed in the cuprates in the low
doping regime \cite{Fuj02}, within the extended Hubbard model 
with next neighbor ($t$) and next-nearest neighbor ($t'$) hopping. 
The latter parameter is expected to play an important role --- 
while the Hartree-Fock (HF) approximation predicts showed that doped 
holes redistribute and the kinetic energy is gained in stripe phases 
for increasing $-t'/t$ \cite{PM'05}, it has been also suggested that 
large $-t'/t$ destabilizes the stripe order.

We implement local electron correlations within a rotationally invariant 
slave-boson (SB) approach in spin space \cite{Fre92}, introducing 
auxiliary boson operators $\{e_i,d_i,p_{i0},\textbf{p}_i\}$ which 
control the actual electronic configuration at each site $i$. The 
Hamiltonian may be then written in the form, 
\begin{equation}
H=-\sum_{ij}\sum_{\sigma\lambda\tau}t_{ij}
   z^{\dag}_{i\sigma\lambda}f^{\dag}_{i\lambda} 
   f^{}_{j\tau}z^{}_{j\tau\sigma} 
  + U\sum_{i}d^{\dag}_{i}d^{}_{i},
\label{eq:Hubb_sb}
\end{equation}
where $\{\underline{z}_i,\underline{z}_j\}$ are $2\times 2$ matrices in 
spin space which depend on the actual configuration of the boson fields. 
When they are replaced by their time-independent averages using 
the mean-field procedure of Kotliar and Ruckenstein \cite{Kot86}, 
the respective amplitudes $t_{ij}$ for the hopping of $f^{\dag}_{j\sigma}$ 
pseudofermions are renormalized. 

We analyzed the stability of various possible stripe phases for the low
doping $x=1/16$. For each stripe phase we performed the symmetry 
analysis, defined its symmetry group, and the relevant magnetic unit
cell. The self-consistent calculations for stripe phases were then 
completed in reciprocal space which allows one to analyze large clusters 
and to eliminate finite-size effects \cite{Rac06}. In order to reach 
fast convergence, the calculations were done at finite but very low 
temperature $\beta t=1000$, with $\beta=1/k_BT$, so that the free energy 
is approximately equal to the internal energy $E$. 

\begin{figure}[t!]
\includegraphics[width=7.4cm]{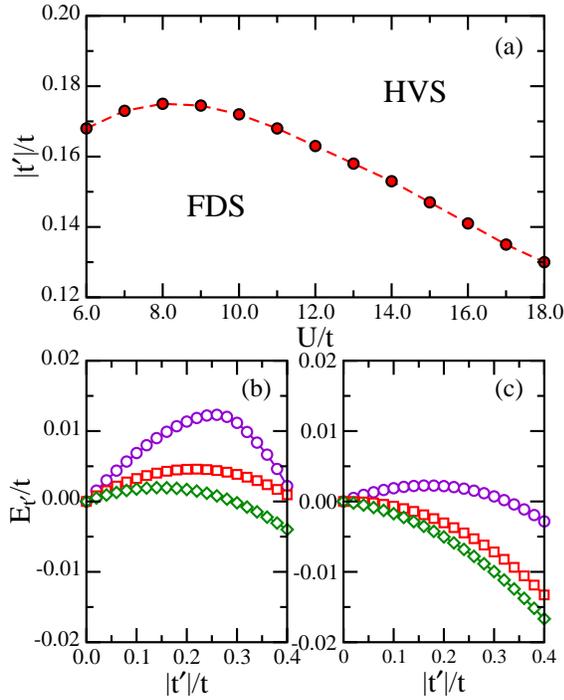}
\caption[]{
(a) Phase boundary between the FDS phase and HVS phase, as obtained for
increasing $U/t$ and next-nearest-neighbor hopping $|t'|/t$.
Average kinetic energy $E_{t'}$ per diagonal bond at the doping $x=1/16$
for increasing $|t'|$ in: 
(b) FDS phase, and (c) HVS phase.
Circles, squares and diamonds show the data obtained for 
$U/t=6$, 12 and 18.
} 
\label{fig:fig1} 
\end{figure}

Taking the relevant parameters for the cuprates with $U=12t$ 
\cite{Fle01}, one finds that the energy of the FDS phase is lower than 
that of the HVS phase (Table~\ref{tab:Ftp}). 
A somewhat larger Coulomb potential 
energy $E_U$ in the FDS phase is accompanied by the large gain of the 
kinetic energy $E_t$ in this phase, in agreement with the solitonic 
mechanism which stabilizes insulating stripe phases, so that 
$\Delta E=E_{\rm FDS}-E_{\rm HVS}=-0.0065t$. At finite $t'=-0.3t$, the 
charge redistributes and the interaction energy $E_U$ is almost almost 
equal in both phases. Although the kinetic energy $E_t$ is still lower 
in the FDS phase, the frustrated next-neighbor kinetic energy $E_t'$ is 
paid in this phase, while it is gained in the HVS phase. As a result,
the HVS phase is more stable and $\Delta E=+0.0051t$.   

\begin{table}[b!]
\caption 
{
Coulomb repulsion energy $E_U$, kinetic energies $E_t$ and $E_t'$, and 
total energy $E$, all per one site, as obtained in the SB approach for 
the FDS and HVS phase for an $128\times 128$ cluster. 
Parameters: $U=12t$ and $x=1/16$.
}
\vskip .1cm
\label{tab:Ftp}
\begin{center}
\begin{tabular}{rcccrc}
\hline
\multicolumn{1}{c}{$t'/t$}     & 
\multicolumn{1}{c}{phase}      & 
\multicolumn{1}{c}{$E_U/t$}    & 
\multicolumn{1}{c}{$E_t/t$}    &
\multicolumn{1}{c}{$E_{t'}/t$} & 
\multicolumn{1}{c}{$E/t$}     \cr
\hline
$ 0.0$ &  FDS  & 0.2903 & $-0.7464$ &   0.0     & $-0.4561$ \cr 
       &  HVS  & 0.2860 & $-0.7356$ &   0.0     & $-0.4496$ \cr    
$-0.3$ &  FDS  & 0.2775 & $-0.7292$ &   0.0039  & $-0.4478$ \cr 
       &  HVS  & 0.2767 & $-0.7225$ & $-0.0071$ & $-0.4529$ \cr    
\hline
\end{tabular}
\end{center}
\end{table}

The crossover from FDS to HVS phase occurs at $|t'|\simeq 0.163t$ at
$U=12t$, and the critical value of $|t'|$ decreases with increasing 
$U$ in the strong coupling regime of $U>8t$, while for smaller values 
of $U$ this trend is reversed (Fig.~\ref{fig:fig1}) in agreement with 
the recent HF results \cite{PM'05}. However, a correct treatment of strong 
electron correlations reduces the critical value of $t'$ nearly twice with 
respect to the HF data. The transition between 
the FDS and HVS phase occurs mainly due to the kinetic energy term $E_{t'}$
which shows a remarkable dependence on $|t'|$, see Figs.~\ref{fig:fig1} (b) 
and \ref{fig:fig1}(c). 
For small $U=6t$ the energy $E_{t'}$ first increases with increasing 
$|t'|$, and starts to decrease only above some finite value 
$t'\sim -0.2t$. Thereby, the kinetic energy $E_{t'}>0$ for the FDS 
phase in a broad range of $U$ and $t'$, while $E_{t'}<0$ for the HVS 
in the strong coupling regime of $U\ge 12t$. As a result, the total 
energy $E$ of the HVS (FDS) phase decreases (increases) with increasing 
$|t'|$.

The above dependence of the kinetic and total energies on the value of
$|t'|$ is generic, and the crossover between the FDS and HVS phase is 
almost independent of the doping. For instance, at
doping $x=1/8$ and for $U=12t$ one finds that the HVS phase
is more stable for $|t'|>0.171t$. This suggests that $|t'|$ increases 
with doping and causes a transition from the FDS to HVS phase at 
doping $x\simeq 1/16$. Hence our findings pose an interesting physical 
problem for future studies --- microscopic derivation of the parameters 
of the extended Hubbard model relevant for the cuprates. 

In summary, we have identified the microscopic origin of the 
reorientation of the domain walls, observed in the cuprates at 
increasing doping. 
By evaluating individual kinetic energy and Coulomb energy 
terms, we established that $t'$ plays an important role in the low 
doping regime, where it controls the crossover from FDS to HVS phase. 
Therefore, in spite of the robust stability of FDS phase at $t'=0$, 
the HVS phase takes over for the realistic value of $t'=-0.3t$ in the
cuprates.


This work was supported by the Polish Ministry of Science and 
Education under Project No. 1~P03B~068~26, and by the Minist\`ere
Fran\c{c}ais des Affaires Etrang\`eres under POLONIUM 09294VH.

\end{document}